# Forty two years counting spots: solar observations by D. E. Hadden during 1890-1931 revisited

V. M. S. Carrasco[1], J. M. Vaquero[1,2*], M. C. Gallego[1] and R. M. Trigo[2,3]

[1]Departamento de Física, Universidad de Extremadura, Spain.
[2]CGUL-IDL, Universidade de Lisboa, Lisbon, Portugal
[3]Departamento de Eng. Civil, Universidade Lusófona, Lisbon, Portugal

**Abstract**: We have recovered the sunspot observations made by David E. Hadden during 1890-1931 from Alta, Iowa. We have digitized the available data published by Hadden in different astronomical journals. This data series have been analyzed and compared with the standard sunspot number series. Moreover, we provide additional information on two great sunspot groups, previously not described, that originated two important extreme episodes of space weather on February 1892 and September 1898.

1. Introduction

Collection of old records is an important task in the context of studies on space climate (Vaquero and Vázquez, 2009). Recovery of old telescopic observations of sunspots has enabled to improve the reliability of long-term reconstructions of solar activity based on the sunspot number (Vaquero, 2007). Most of these studies rely on two sunspot numbers series widely available. The two sunspot series are discussed, e.g., in a review by Usoskin (2013). On the one hand, the classical Wolf Number, that is currently updated by the Solar Influences Data Analysis Center that produces the International Sunspot Number *ISN* (Clette et al., 2007). On the other hand, many studies use the more recently derived Group Sunspot Number *GSN* developed by Hoyt and Schatten (1998) (hereafter HS98).

Recently, Svalgaard (2012) has proposed that the sunspot number series can be affected by problems of homogeneity. This has underlined the interest of recovering long-term series of observations made by the same observer in order to obtain homogeneous series without problems. It is within this context that the main aim of this

* Corresponding author: J. M. Vaquero (jvaquero@unex.es)

Centro Universitario de Mérida,

Avda. Santa Teresa de Jornet, 38

06800 Mérida (Badajoz), Spain

Tel.:+34 924 289300

Fax: +34 924 301 212

paper is to retrieve and analyze the 42 year long sunspot observations made by D. E. Hadden in late 19th century and early 20th century, which have barely been considered until now.

Before describing the data currently available we offer some biographical notes of D. E. Hadden. These data have been recovered by consulting the original sources published in several scientific journals. We also analyzed these data and compared with other indices of solar activity based on sunspot counting. Finally, we analyze in some detail the observations by Hadden relative to two interesting cases for studies of extreme space weather events.

## 2. Biographical notes of D. E. Hadden

David E. Hadden was born in Ireland on 22 October 1866. In 1881, his family immigrated to the United States where, after a short stay in Le Mars (Iowa), moved again to South Dakota where his father exerted his profession as a doctor. They remained there for the following twelve years until they settled in Alta, Iowa (Harlan, 1931).

Hadden studied for two years at Wesley College in Dublin before coming to the United States. In 1904, he graduated from the Bachelor of Science in Chemical and Pharmacy, at Morningside College in Sioux City where he had enrolled in 1903. He was appointed to the State Pharmacy Board in 1909.

He got married in September 1889 with Emeline Dier of Le Mars (Iowa) and was the father of Lola E. and Edward A. His hobbies were astronomy and meteorology (Fig. 1). So, he built an excellent small observatory at home (Hadden, 1910). His work was highly regarded, excelling in the field of sunspots. He became a member of the Iowa Academy of Science, American Astronomical Society and British Astronomical Association. He began his meteorological observations on January 1890 and made daily readings during 53 years. Finally, David E. Hadden died on September 20, 1943 at the age of 77 years old (Anonymous, 1943).

## 3. Available data

In this work, we have collected all the data published by D. E. Hadden on solar observations. However, these data are scattered in numerous articles published in several astronomy or academic journals, partially explaining why its use has been so restricted. Recovered data have been obtained from the following journals: *Memoirs of the British Astronomical Association* (1900-1906), *Monthly Review of Iowa Weather and Crop Service* (1890-1897), *Popular Astronomy* (1895-1931), *Proceedings of the Iowa Academy of Science* (1894-1922) and *Publications of the Astronomical Society of the Pacific* (1893-1897). Table 1 concisely shows what data are available on the number of sunspot groups, number of individual spots, number of facular regions, and the time scale of these numbers (daily or monthly).

We must emphasize that the first observations made by Hadden between 1890 and 1896 were obtained with a 3-inch refractor telescope (Hadden, 1934). He replaced this first telescope by a Brashear 4-inch refractor telescope that he used during the period 1897-1906. Also, he used an 8½-inch reflector telescope simultaneously with the Brashear refractor between 1903 and 1905 (Hadden, 1906, 1905, 1897). Beginning in 1907, Hadden made most of his subsequent observations with a 5½-inch refractor telescope and, occasionally, with a Brashear 9½-inch reflector (Hadden, 1914). Unfortunately, during these periods where Hadden had two telescopes available, it is impossible to be sure of which instrument he used to perform the recorded observation at any single day. Finally, Hadden had one small spectroscope in his observatory incorporating an excellent 2-inch Rowland diffraction grating. He usually made observations with this instrument (Hadden, 1904).

Annual data obtained from the observations made by Hadden are listed in Table 2 where $N$ is the number of days observed, $G$ is the annual average number of sunspot groups, $S$ is the annual average number of spots and $F$ is the annual average number of faculae. Following the usual definitions of sunspot number (HS98; Clette et al., 2007), we can define a Group Sunspot Number from the values of $G$ obtained by Hadden ($GSN_H$) and a Wolf or International Sunspot Number from the values of $G$ and $S$ obtained by Hadden ($ISN_H$) using the expressions $GSN_H = 12.08G$ and $ISN_H = 10G + S$ respectively.

Using the above mentioned expressions we have derived the monthly and annual indices reconstructed from the papers by Hadden; $GSN_H$, $ISN_H$ and number of facular regions (Fig. 2). Additionally, the bottom panel shows the monthly number of days with records. We should note that the number of faculae $F$ (Fig. 2C), presents a striking peak (at the monthly scale), close to the maximum of solar cycle 16 and far above all other values. This is due to the fact that there are only two days of observation in that month.

The vast majority of data published by Hadden in his papers corresponds to monthly resolution. We have recovered monthly data from 1890 to 1931. To the best of our knowledge, no solar record exists after 1931. For the periods 1917-1920 and 1930-1931, we found only values of $G$ and, therefore, the values for $S$ and $F$ are lost. Monthly data retrieved correspond to solar cycles 13-16. In general, the values obtained by Hadden look similar to the appearance of the official data of the $GSN$ and $ISN$.

With regard to daily data, it has been recovered from April 1890 to December 1902, except for the period spanning between May 1893 and May 1894. These daily data correspond to the solar cycle 13 (whose maximum occurred in 1893-1894) and the beginning of solar cycle 14. These data are available as an electronic supplement to this paper. It is worthy of mention that the daily data from Hadden that HS98 included in their reconstruction of solar activity are minimal, corresponding only to new groups that appeared in the solar disc during three months, from October to December 1890 (Hadden, 1891). Moreover, we have recovered daily data from all the additional sources mentioned by HS98 that provide daily values for $G$, $S$ and $F$ for a wider period. Therefore, we can state without any doubt that the new data for sunspot group constructed here fully replaces the dataset provided by HS98, increasing the number of records significantly. It is striking that the number of annual records obtained from these daily observations is greater than the corresponding number obtained from the annual tables provided by Hadden (Table 2). Our hypothesis is that Hadden published a high number of days with zero spots during solar minimum of 1900-1902 that correspond merely to interpolated data between two dates with no spots detected.

4. Analysis

In the previous section, we have presented the general characteristics of the monthly and annual solar activity indices that have been collected. In this section, we will develop an analysis of these series, namely assessing their variability and reliability. We should start by acknowledging that there are advantages and disadvantages in this data set. The main advantage is that it has recovered a high number of data from a single observer with over 40 years of observation, including four solar cycles (SC13-SC16). Furthermore, we were able to retrieve a non-common parameter in historical records of solar data as the number of faculae (besides sunspots and groups counting). Finally an additional advantage is the internal coherence and homogeneity of the entire dataset, i.e. that there are no changes in the criteria of the observer for counting spots, groups, and faculae. Nevertheless, this data set presents a clear disadvantage. Although we know the telescopes used by the observer, we do not have details about eventual changes of telescope for periods where Hadden had more than just one instrument available. We also know that during very short periods of absence of Hadden, her daughter was in charge of continuing the observations (Anonymous, 1943). However, the sources are not detailed in respect to such periods namely who performed the observations and what was observed and cataloged during these short absences.

In this section we analyze the sunspot numbers obtained by Hadden ($ISN_H$ and $GSN_H$) with the following two main objectives; (i) comparing the data from their observations to other indices and (ii) studying the comparison between different data provided by the observer.

We will compare the values obtained by Hadden with the indices $ISN$ and $GSN$. For this purpose we have calculated the annual ratios $ISN_H/ISN$ and $GSN_H/GSN$ and the evolution of these values is shown in Fig. 3. In this figure squares (circles) represent the $ISN_H/ISN$ ($GSN_H/GSN$) ratio and solid (dashed) lines show the mean value of $ISN_H/ISN$ ($GSN_H/GSN$) ratio before and after of 1903. There is an abrupt change in these values just in 1903 (Fig. 3) when Hadden started to use an 8 ½-inch reflector telescope (1903-1905). We may note that he had previously two refractor telescopes (4-inch and 3-inch). Therefore, this change in amplification magnitude was remarkable. From 1907 onwards, Hadden used a 5 ½-inch refractor, however sometimes complemented by the use of the 9 ½-inch reflector. Unfortunately, we do not know when he made use of one or the other instrument. Note that higher values of the ratios after 1903 are related with low values of sunspot number in solar cycle minima.

We can also compare the annual values of *ISN* and *ISN$_H$* in order to compute the calibration constant for the observer (Fig. 4). The expression for the best linear fit is the following: $ISN = (0.89 \pm 0.03) \cdot ISN_H + (0.8 \pm 1.5), r = 0.982$, *p*-value $< 0.001$. However, if we set the y-intercept of the line equal to zero, we can obtain a calibration constant $k_H = 0.91 \pm 0.02$ for D. E. Hadden.

An interesting index to evaluate the performance of the derived *ISN$_H$* is the *ISN$_H$/G* ratio. Hoyt et al. (1994) showed that this ratio is close to 12 taking into account the average of modern observers and in Fig. 5 we show the relationship of the data obtained for Hadden. The best linear fit obtained between these series is given by $ISN_H = (15.3 \pm 0.2) \cdot G + (-1.0 \pm 0.6), r = 0.997$, *p*-value $< 0.001$. If we set the y-intercept of the line equal to zero, we obtain $ISN_H / G = 15.0 \pm 0.1$. This value is somewhat higher than the value estimated by Hoyt et al. (1994) for the average of modern observers. As a comparative example, we can cite the value of $13.4 \pm 0.5$ obtained for the Portuguese astronomer Melo e Sima who observed during 1895-6 (Vaquero et al., 2012).

We will represent the ratio between *GSN$_H$* and *ISN$_H$* calculated with annual data to study the combined variability of these indices. We observe no significant change in this ratio (Fig. 6). Note that the error bars in Fig. 6 represent one standard deviation of this dataset. Moreover, we show in Fig. 7 the ratio of the number of groups and the number of faculae (*G/F*) for annual data obtained with the data registered by Hadden. Although the series does not present a regular behaviour, such as the solar activity cycle, this ratio has a performance similar to other solar indexes (i.e., sunspot number). We can see that the maxima of this ratio are located in the years 1893, 1906, and 1928 corresponding to the maxima of solar cycles 13, 14, and 16 respectively. Note that data for the maximum of solar cycle 15 the ratio are not available due to the lack of *F* values for several years (Table 1). Minima values for the *G/F* ratio are located in the years 1899, 1912, and1921. Whereas the maxima for the *GSN$_H$* are located in the years 1893, 1905, and 1928, the minima are located in the years 1901, 1913 and 1923. In the case of *GSN*, the maxima are located in the years 1894, 1907 and 1928. We should note that the number of facular regions (*F*) has a singular behaviour: the peaks corresponding to *F* are reached before the occurrence of the highest values relative to *S* and *G*.

## 5. Great sunspots observed by D. E. Hadden

### 5.1. Great sunspot of February 1892

One of the largest sunspot groups observed by Hadden occurred in early February 1892 (Hadden, 1892). The group appeared on February 5 by rotation in the south hemisphere. Hadden does not observe it again until the February 8 due to bad weather in Alta. He appreciated large activity in the group due to the formation of new spots along its transit of the solar disk. In the last days of the transit, Hadden describes how the penumbra area decreases somewhat. The group disappeared by the solar limb on February 18. Interestingly, Hadden reported that this group was visible by naked-eye using a smoked or shading glass. Hadden (1892) presented a drawing (Fig. 8) showing the evolution of the sunspot group during the days 5-17 of February.

It is worthy to mention that this great spot and its corresponding geomagnetic storm have been described in the well-known popular science book "The Heavens and their Story" by Annie S. D. Maunder and E. Walter Maunder. In fact, photographic trace of the variations of the geomagnetic declination and the horizontal component of geomagnetic field (Fig. 9) was included in Chapter 10 of this book (Maunder and Maunder, 1908, plate XXXVI). The work made these authors provides a nice example of the efforts made by scientists at the time (such as the Maunders) to establish meaningful relationships between solar variability and the increasingly available measurements of terrestrial magnetism (Readers interested in the full description of the work by the Maunders are directed to Soon and Yaskell, 2003).

According to Jones (1955) in the 'Great Geomagnetic Storms recorded at Greenwich-Abinger, 1874-1954', this great spot group (listed with the reference number 2421) corresponds to the solar source of the great geomagnetic storm occurred in 1892, February 13. Jones (1955) lists this storm with the reference number 17. In this storm, the variations of the geomagnetic field components were >73' for the geomagnetic declination (D), >540γ for the horizontal component (H), and >635γ for the component Z. This geomagnetic storm is ranked 23 in the ranking of storms developed by NOAA (http://www.ngdc.noaa.gov/stp/geomag/aastar.html) using the geomagnetic index aa*. Therefore, it is by several standards one of the largest solar storms that have been registered in the last 140 years.

### 5.2. Great sunspot of September 1898

Hadden (1899) provides some interesting information on the great sunspot that appeared by rotation at the Eastern limb of the Sun on the morning of September 2 1898. Due to weather conditions, Hadden did not use his spectroscope until September 6 when no disturbance was observed around this sunspot. The following day, Hadden started the spectroscopic observation at 11.40 am (central time) observing a sudden outburst. Hadden (1898) describes in this form the phenomenon: "[…] *the entire region just preceding, and for some distance following, and also north and south of the spot, was greatly agitated; the Hα line being reversed and distorted; small black jets projecting from either side of the line were noted in several places, and on opening the slit slightly, the flame and spike-like form of the disturbances could be clearly seen. At 12 m. intensely brilliant flames were observed over the large spot extending from the umbra to the edge of the penumbra on the east side. This phenomenon was particularly striking; the intensely bright scarlet flame nearly in the center of the dark absorption band of the spot spectrum, being very interesting. The $D_3$ line was bright; while $D_1$ and $D_2$ and many other lines were widened. At 12h. 5m. p.m a small dark line attached to the Hα line extended obliquely towards the red end of the spectrum, in the region just preceding the main spot. Observations were interrupted at 12h. 10m., and could not be resumed until 1h. 40m., but the entire disturbance had then almost ceased, only a few* *slight reversals being noted. A 2-inch diffraction grating, attached to a 4-inch telescope, was used"*.

According to Jones (1955) in the 'Great Geomagnetic Storms recorded at Greenwich-Abinger, 1874-1954', this great spot group (listed with the reference number 4781) is the solar source of the great geomagnetic storm occurred in 1898, September 9. This storm was listed with the reference number 36 by Jones (1955). The variations in the geomagnetic field components for this storm were respectively, 57' for D, >365γ for H, and 480γ for the component Z. This storm is ranked 71 in the ranking of storms developed by NOAA (http://www.ngdc.noaa.gov/stp/geomag/aastar.html) using the geomagnetic index aa*, being also one of the largest solar storms that have been recorded since 1868.

6. **Conclusions**

We have made a compilation of all data available in the scientific literature of solar observations made by David E. Hadden during the period of 1890-1931, including number of sunspot groups, spots and number of facular regions. The vast majority of these observations were not previously compiled in the database by HS98. We have managed to recover daily data for the period 1890-1902. However, only monthly data are available for the remaining period. We have made a statistical analysis of $ISN_H$ and $GSN_H$ indices computed from Hadden data. In general, these values are comparable to standard values of ISN and GSN. Besides obtaining the calibration constant for Hadden as observer, we have calculated the $ISN_H/G$ ratio detecting no significant changes. Finally, we have recovered some detailed information of two large sunspot groups that caused two of the major magnetic storms recorded in the last 140 years.

**Acknowledgements**

We wish to acknowledge the assistance provided by Craig Johnson (Executive Director, Iowa Academy of Science), The Iowa State University Library and The University of Iowa Libraries. J.M. Vaquero has benefited from the impetus and participation in the Sunspot Number Workshops (http://ssnworkshop.wikia.com/wiki/Home). Support from the Junta de Extremadura (Research Group Grant No. GR10131) and Ministerio de Economía y Competitividad of the Spanish Government (AYA2011-25945) is gratefully acknowledged.

**TABLE 1**

Data retrieved from the observations made by D. E. Hadden on the number of sunspots *S*, groups *G* and faculae *F* between 1890 and 1931 (A indicates availability of daily data and B indicates that only monthly data is available).

|          | 1890 | 1891 | 1892 | 1893 | 1894 | 1895 | 1896 | 1897 | 1898 | 1899 | 1900 | 1901 | 1902 | 1903 |
|----------|------|------|------|------|------|------|------|------|------|------|------|------|------|------|
| **SUNSPOTS** | A | A | A | A | A | A | A | A | B | A | A | A | A | B |
| **GROUPS**   | A | A | A | A | A | A | A | A | A | A | A | A | A | B |
| **FACULAE**  | A | A | A | A | A | A | A | A | B | A | A | A | A | B |
|          | 1904 | 1905 | 1906 | 1907 | 1908 | 1909 | 1910 | 1911 | 1912 | 1913 | 1914 | 1915 | 1916 | 1917 |
| **SUNSPOTS** | B | B | B | B | B | B | B | B | B | B | B | B | B | - |
| **GROUPS**   | B | B | B | B | B | B | B | B | B | B | B | B | B | B |
| **FACULAE**  | B | B | B | B | B | B | B | B | B | B | B | B | B | - |
|          | 1918 | 1919 | 1920 | 1921 | 1922 | 1923 | 1924 | 1925 | 1926 | 1927 | 1928 | 1929 | 1930 | 1931 |
| **SUNSPOTS** | - | - | - | B | B | B | B | B | B | B | B | B | - | - |
| **GROUPS**   | B | B | B | B | B | B | B | B | B | B | B | B | B | B |
| **FACULAE**  | - | - | - | B | B | B | B | B | B | B | B | B | - | - |

**TABLE 2**

Annual data retrieved from the observations made by D. E. Hadden. *N* indicates the number of days observed, *G* the average number of groups, *S* the average number of spots and *F* the average number of faculae.

| YEAR | N | G | S | F |
|---|---|---|---|---|
| **1891** | 257 | 2.9 | 14.9 | 3.6 |
| **1892** | 205 | 5.6 | 34.8 | 4.1 |
| **1893** | 177 | 6.6 | 36.6 | 4.0 |
| **1894** | 129 | 5.6 | 27.5 | 3.4 |
| **1895** | 149 | 5.2 | 30.5 | 3.5 |
| **1896** | 197 | 3.3 | 17.8 | 2.9 |
| **1897** | 198 | 2.4 | 11.9 | 2.4 |
| **1898** | 234 | 2.1 | 11.0 | 2.4 |
| **1899** | 259 | 1.1 | 4.8 | 1.5 |
| **1900** | 255 | 0.7 | 3.4 | 1.0 |
| **1901** | 269 | 0.3 | 0.9 | 0.3 |
| **1902** | 230 | 0.4 | 2.0 | 0.5 |
| **1903** | 230 | 1.7 | 7.9 | 1.9 |
| **1904** | 172 | 3.1 | 13.3 | 3.1 |
| **1905** | 202 | 4.4 | 18.2 | 3.9 |
| **1906** | 198 | 3.8 | 15.0 | 3.3 |
| **1907** | 205 | 3.8 | 15.8 | 3.4 |
| **1908** | 231 | 3.4 | 13.0 | 3.4 |
| **1909** | 121 | 3.0 | 12.8 | 3.0 |
| **1910** | 171 | 1.3 | 5.0 | 1.9 |
| **1911** | 156 | 0.6 | 1.6 | 1.0 |
| **1912** | 150 | 0.4 | 1.8 | 0.7 |
| **1913** | 159 | 0.1 | 0.6 | 0.3 |
| **1914** | - | 1.0 | 3.5 | 1,4 |
| **1915** | - | 3.0 | 16.4 | 3.7 |
| **1916** | - | 4.5 | 21.7 | 3.9 |
| **1917** | - | 7.3 | - | - |
| **1918** | - | 5.7 | - | - |
| **1919** | - | 4.5 | - | - |
| **1920** | - | 2.4 | - | - |
| **1921** | 161 | 1.8 | 8.4 | 2.3 |
| **1922** | 161 | 1.1 | 5.0 | 1.2 |
| **1923** | 173 | 0.6 | 2.4 | 0.9 |
| **1924** | 172 | 1.2 | 7.1 | 1.6 |
| **1925** | 115 | 3.1 | 17.3 | 3.1 |
| **1926** | 100 | 4.4 | 20.7 | 4.1 |
| **1927** | 75 | 4.7 | 25.6 | 4.1 |
| **1928** | 109 | 5.4 | 28.7 | 3.8 |
| **1929** | 49 | 4.7 | 22.9 | 3.8 |
| **1930** | 115 | 2.7 | - | - |
| **1931** | 139 | 1.6 | - | - |

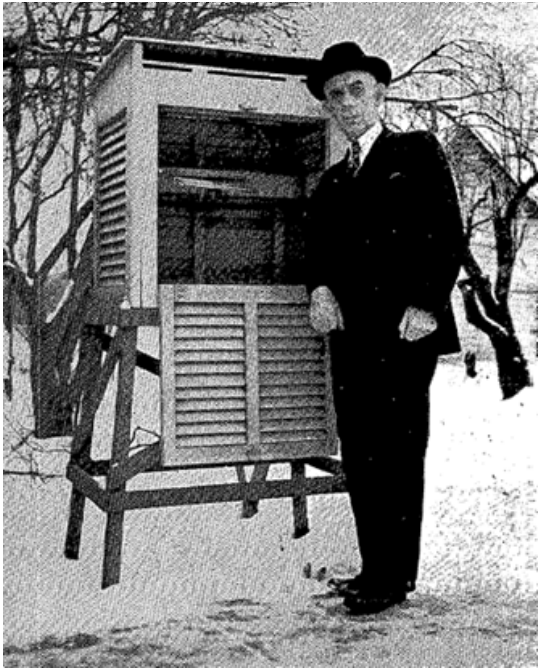 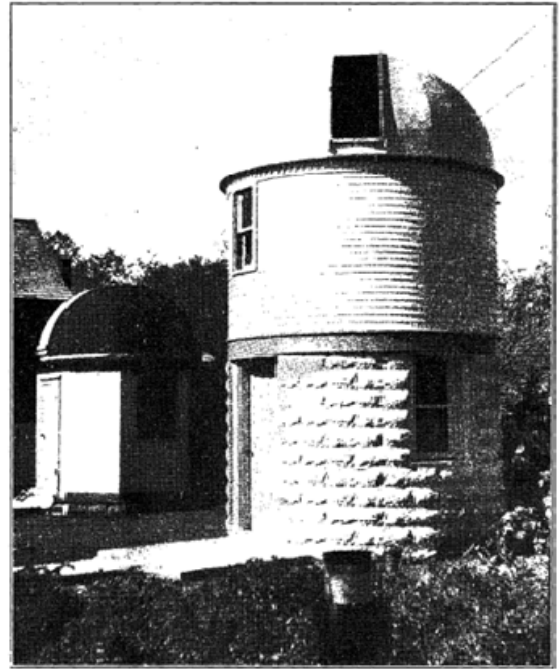

**Fig. 1.** Mr. Hadden and his thermometer shelter (left) and his astronomical observatory (right). [Source: *Anonymous, 1943. David E. Hadden broke record for long service. Climatological Data. U. S. Department of Commerce, Weather Bureau, LIV, 120* and *Hadden, David E, 1910. An amateur´s observatory. Popular Astronomy 18, 597-600.*].

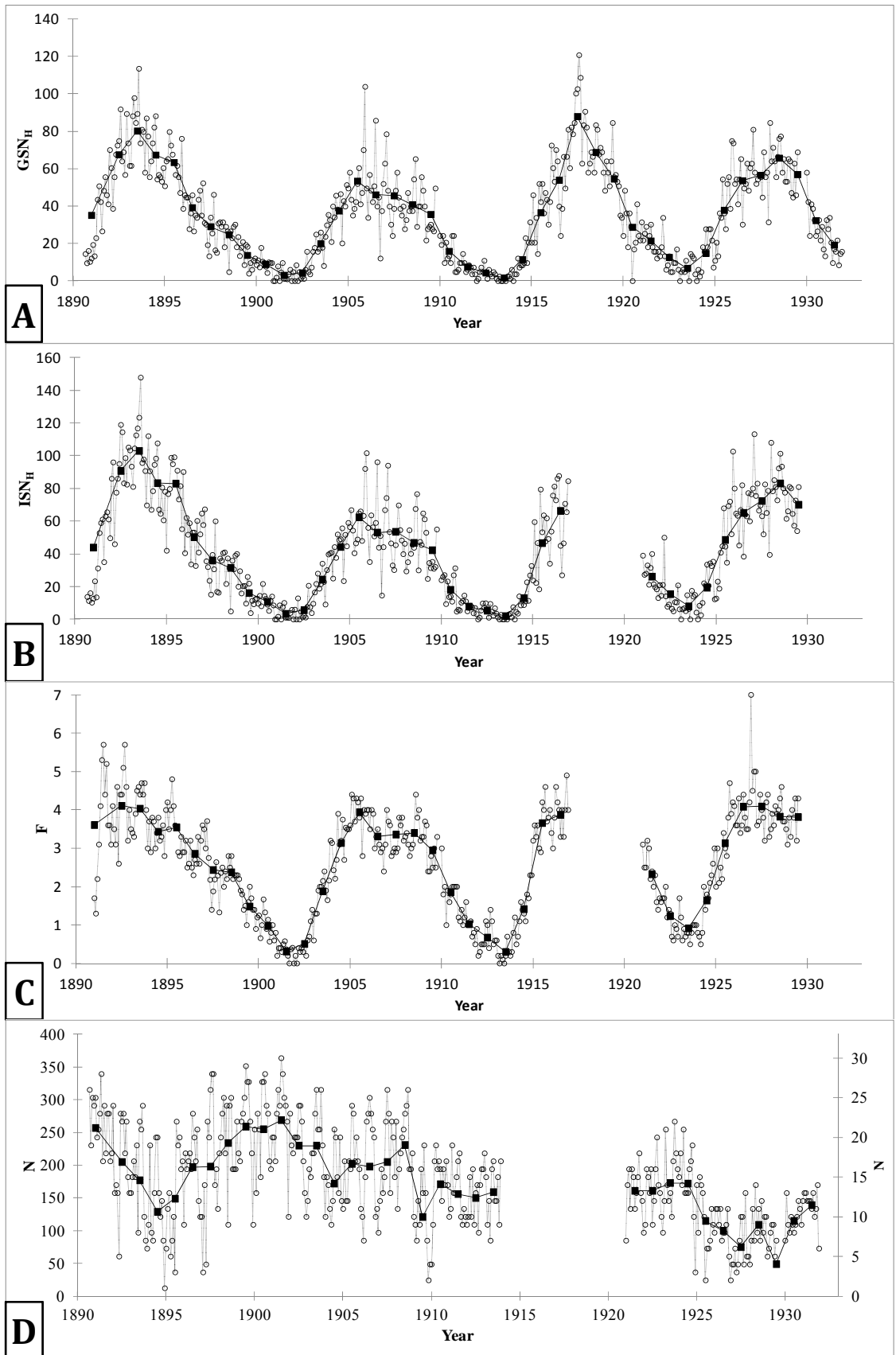

**Fig. 2.** Monthly (circle) and annual (square) data from observations of Hadden: (*A*) $GSN_H$, (*B*) $ISN_H$, (*C*) number of faculae *F*, and (*D*) number of days with observation *N*. The y-axis for annual (monthly) values is located on the left (right) of the panels.

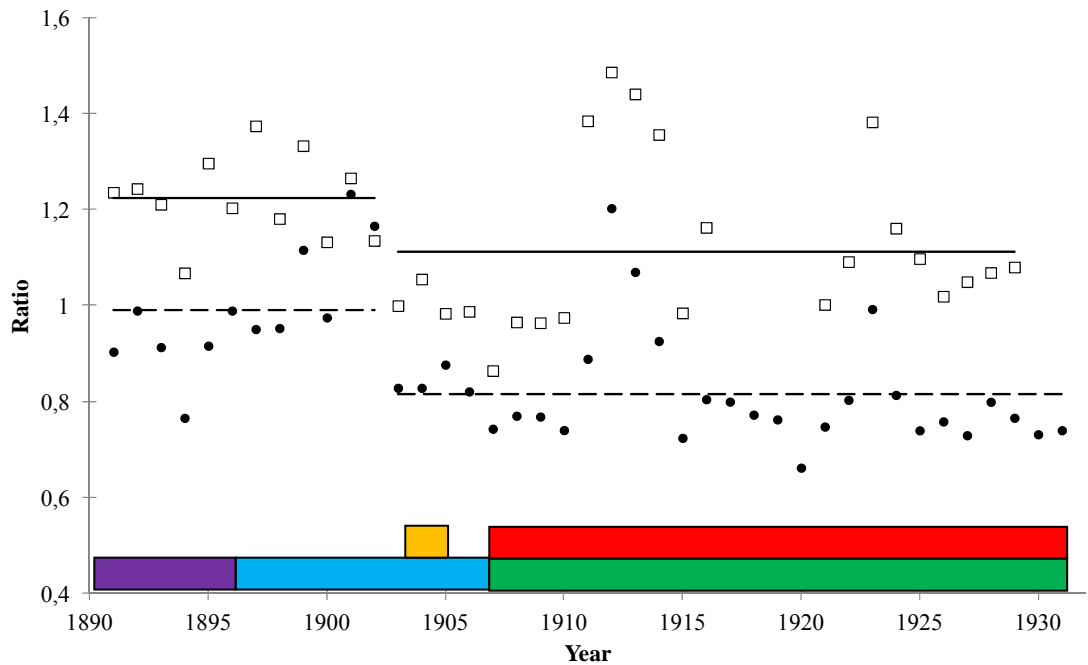

**Fig. 3.** Time evolution of annual ratios $ISN_H/ISN$ (squares) and $GSN_H/GSN$ (circles). Color bars indicate the telescope used by Hadden: (purple) 3-inch refractor telescope, (blue) Brashear 4-inch refractor telescope, (orange) 8½-inch reflector telescope, (green) 5½-inch refractor telescope and (red) Brashear 9½-inch reflector telescope. Continuous and dashed lines represent the mean values of annual ratios $ISN_H/ISN$ and $GSN_H/GSN$, respectively, for 1890-1902 and 1903-1931.

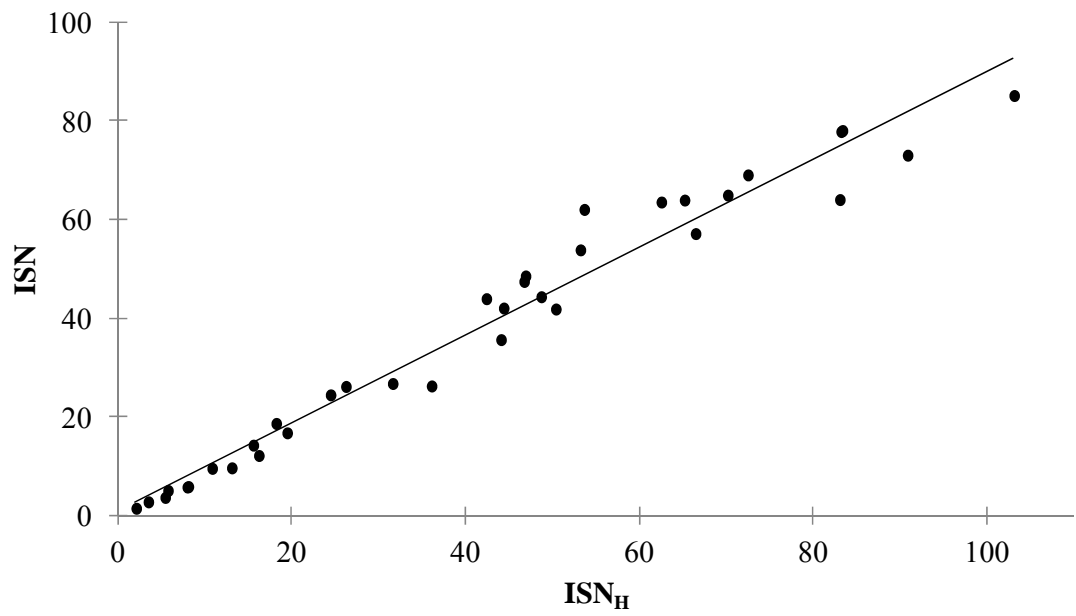

**Fig. 4.** Linear relationship between *ISN$_H$* and *ISN*.

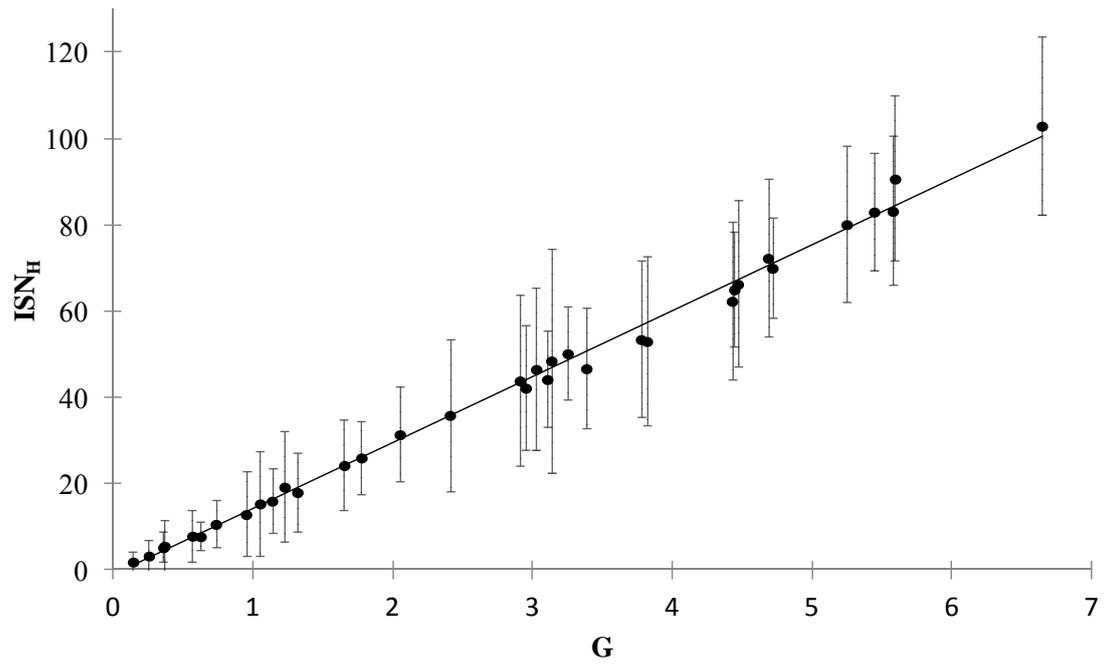

**Fig. 5.** Relationship between annual $ISN_H$ and the number of groups observed by Hadden $G$. Error bars represent one standard deviation.

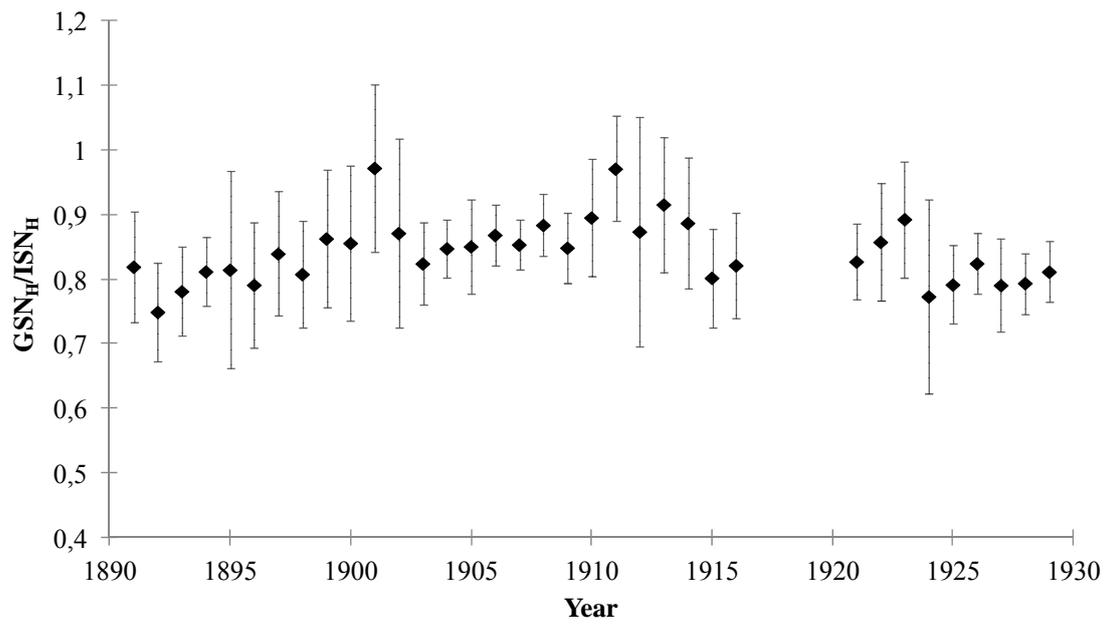

**Fig. 6.** Ratio of annual values $GSN_H$ e $ISN_H$ with error bars (one standard deviation).

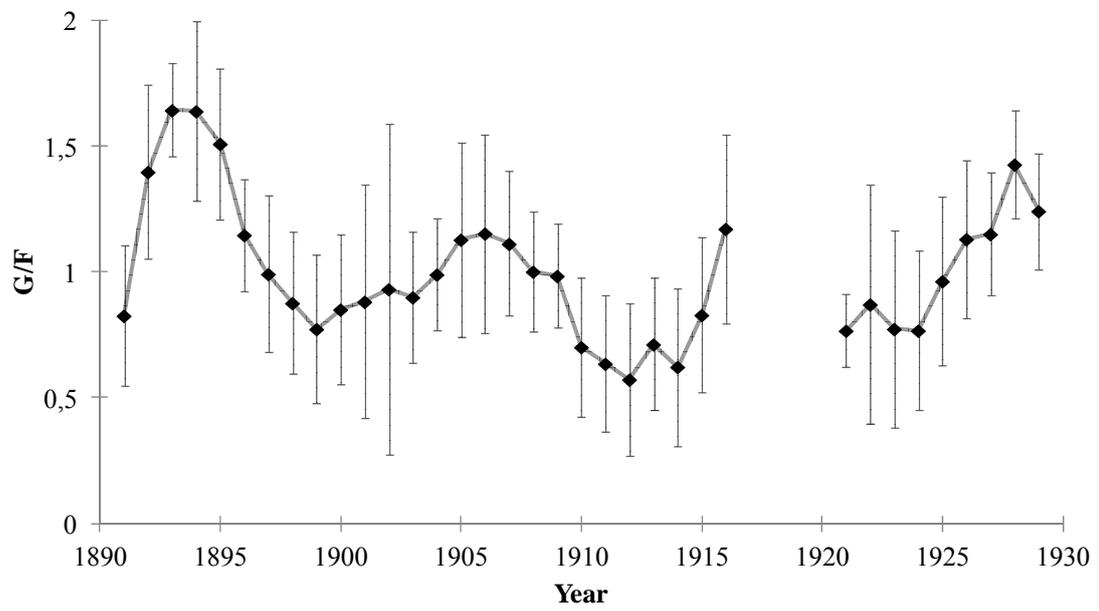

**Fig. 7.** Ratio between the number of annual group *G* and the annual number of faculae *F* (error bars indicate one standard deviation) obtained by Hadden.

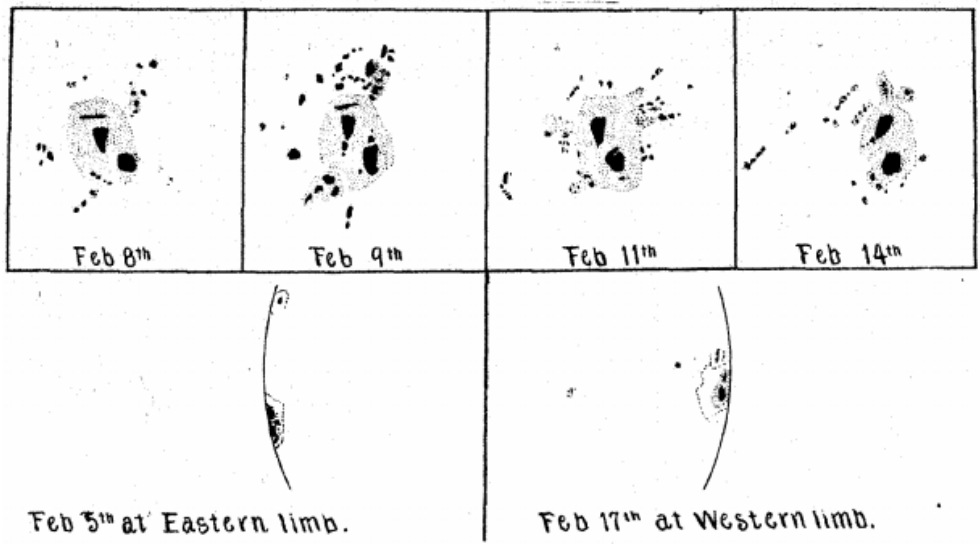

**Fig. 8.** Great sunspot of February 1892 [Source: *Monthly Review of Iowa Weather and Crop Service III, 9*].

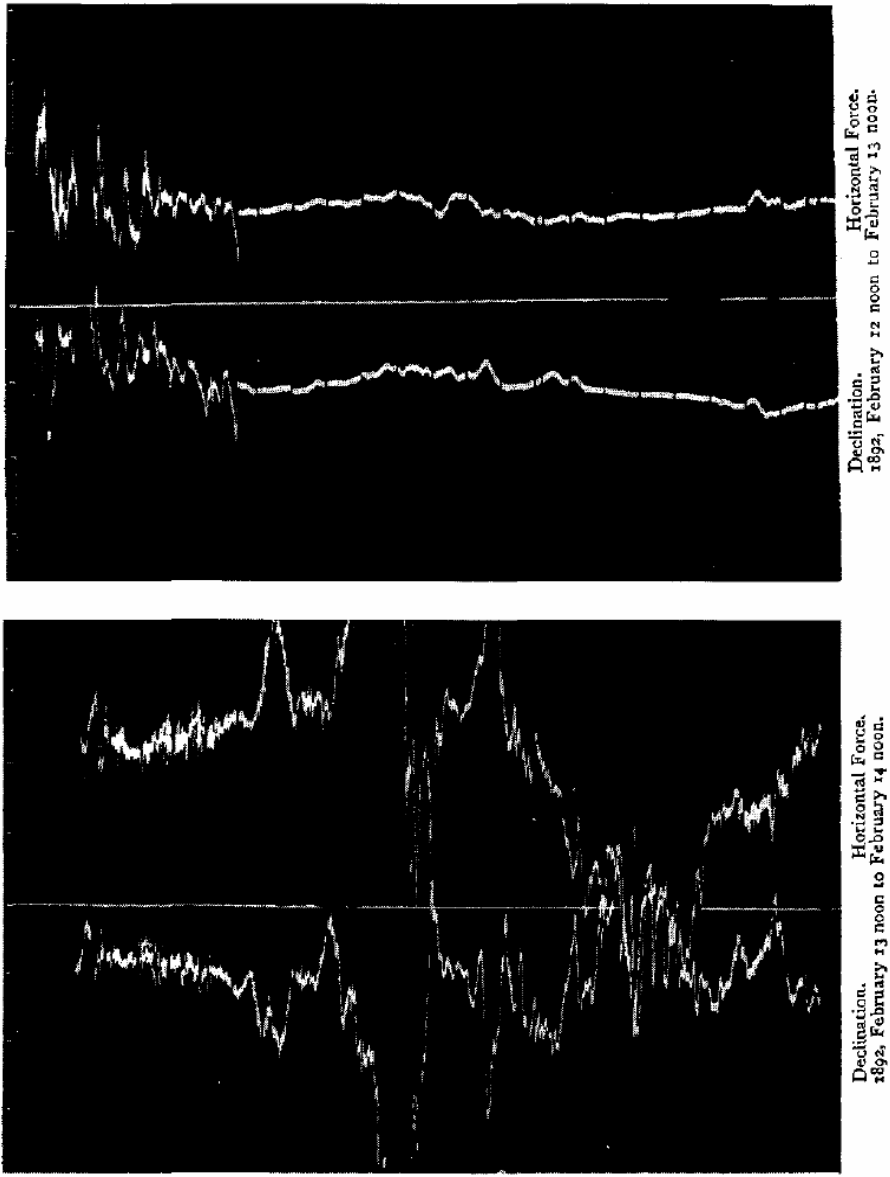

**Fig. 9.** Magnetograms of the storm of February 13, 1892 published by Maunder and Maunder (1908) from the photographic registers taken at the Royal Greenwich Observatory (reprinted from Maunder and Maunder, 1908, p. 182, plate XXXVI).